\documentclass[aps,pra,letterpaper,twocolumn,showpacs,amssymb,amsmath,amsfonts,superscriptaddress,floatfix,nofootinbib]{revtex4}
\usepackage{graphicx,graphics,times,bm,bbm,multirow,color}
\newcommand{\be}{\begin{eqnarray}}
\newcommand{\ee}{\end{eqnarray}}

\begin{document}

\title{Temperature effects on quantum cloning of states and entanglement}

\author{S. Baghbanzadeh}
\affiliation{Department of Physics, Sharif University of Technology, P. O. Box 11155-9161, Tehran, Iran}
\affiliation{Department of Physics, Iran University of Science and Technology, Narmak, P. O. Box 16765-163, Tehran, Iran}
\author{A. T. Rezakhani}
\affiliation{Center for Quantum Information Science and Technology, and Departments of Chemistry and Physics,\\ University of Southern California, Los Angeles, CA 90089, USA}
\affiliation{Institute for Quantum Information Science, University of Calgary, Alberta T2N 1N4, Canada}

\begin{abstract}
Performances of the symmetric universal and phase-covariant cloning transformations and entanglement cloners --- qubit case --- are investigated when the initial state of the hardware or the original state to be cloned is weakly coupled to a thermal environment. Different behaviors of each of these transformations are analyzed and contrasted with the ideal cases.

\end{abstract}

\maketitle

\emph{Introduction.}--- It has been known that based on the principles of quantum mechanics, accurate cloning of arbitrary quantum state is impossible \cite{WZ82}. Nevertheless, on the account of the significant role of copying in quantum computation and quantum communication, a variety of approximate quantum cloning transformations have been proposed, e.g., universal cloning (UC) machines producing identical copies from arbitrary input qudits \cite{UC0,UC1,UC2,UC3,UC4}, phase-covariant cloning (PCC) machines of a class of partially known qudits \cite{PCC1,PCC2}, and optimal entanglement cloning machines \cite{EC,Locc}. For recent reviews see Refs.~\cite{review}.

A question of practical relevance is how uncontrollable environmentally induced decoherence or dissipation can affect performance of quantum cloning machines. In closed systems, an initially pure state evolves to another pure state. In practice, however, preparation of pure states and/or keeping them pure are not generally easy tasks. In general, interaction with an environment degrades purity of quantum systems and makes their states mixed. A usual effect that a thermal environment can cause is thermalization (as a kind of dissipation)~\cite{dectherm,therm}. That is, because of interaction with the environmental degrees of freedom which are in thermal equilibrium, the quantum system will also be driven toward equilibrium. It should be noted that a generic isolated quantum many-body system does also relax to a state well described by the standard statistical-mechanical prescription \cite{iso}. In this paper, our aim is to investigate temperature effects on the performance of the cloning machines. It has been known that decoherence can put a limitation on the number of clones that a quantum cloner can generate \cite{limit}. There is also a model in which the robustness of the cloner increases with the number of qubits \cite{ZG05}.

Through a thermalizing process, the density matrix of the system $\varrho$ in long time will approach the Boltzmann state $\varrho_{\text{th}}=e^{-\beta H}/Z$, where $k_B\beta$ is inverse temperature ($k_B$ is the Boltzmann constant), $H$ is the Hamiltonian of the open quantum system, and $Z=\text{Tr}(e^{-\beta H})$ is the partition function. Energy and phase relaxation processes, with the time-scales $T_1$ and $T_2$, respectively, are common processes present when approaching an induced equilibrium state. For a more precise and elaborate discussion of dissipation, thermalization, and decoherence see Ref.~\cite{dectherm}. Some (phenomenological) models for the underlying \textit{dynamics} of the thermalization have already been proposed \cite{therm}. We assume that the time-scale in which typical correlation functions of the environment decay, $t_\text{env.}$, is much smaller than all other time-scales, i.e., $t_\text{env.}\ll\tau_\text{c},T_\text{diss.}=\min\{T_1,T_2,T_\text{O}\}$, where $\tau_\text{c}$ is the time-scale of the cloning process and $T_\text{O}$ is the time-scale dictated by all other relaxation mechanisms. This assumption is important for the Markovian analysis of the dynamics of the thermalization \cite{dynamics,BP}. This implies that during the cloning process, a negligible amount of information flows from the environment to the system (or vice versa). Here, we also assume that $\tau_\text{c}\lesssim T_\text {diss.}$. This extra condition allows us to ignore dynamical effects of the thermalization, hence consider a simple \textit{static} (toy) model --- explained below --- to bring temperature into play. Despite these simplifying assumptions, we will argue that the result is still reliable enough to give a hint about how temperature effects can change performance of different cloning machines such as the universal cloners, phase-covariant cloners, and entanglement cloners. Indeed, such investigation has an immediate importance in attempts to realize quantum cloning in systems where (due to thermal and perhaps other noise effects) the preparation of pure states, whether initial state of the system to be cloned or the quantum hardware, is difficult, such as in NMR systems \cite{NMR1,NMR2}. For another study using a different approach, see Refs.~\cite{LCDT,starnet}. For the purpose of illustration, we only consider the case of symmetric $1\rightarrow 2$ qubit cloners. Extension to $M\to N$ qudits is straightforward as well.

\emph{Optimal universal and phase-covariant cloning transformations.}--- In the universal cloning transformation, it is usually assumed that the qubit state to be cloned is a pure state, $|\Psi\rangle_a=\cos\frac{\theta}{2}|0\rangle+e^{i\phi}\sin\frac{\theta}{2}|1\rangle$, and the blank copy ($b$) and the quantum cloning machine (also called ancillary system, $c$) are each in a known pure state, say $|0\rangle$ \cite{UC0,UC1,UC2,blank}. The symmetric cloning transformation, then, acts in this way: $U\left(|\Psi\rangle_a|0\rangle_b|0\rangle_c\right)=|\Upsilon\rangle_{abc}$, where $\text{Tr}_{bc}(|\Upsilon\rangle_{abc}\langle\Upsilon|)=\text{Tr}_{ac}(|\Upsilon\rangle_{abc}\langle\Upsilon|)$. The latter condition guarantees that the final clones both have the same states, $\varrho^\text{out}_a=\varrho^\text{out}_b$. A measure to quantify performance of a cloning machine is the fidelity between the original and the output states,  $F(\rho,\sigma)=\left(\text{Tr}\sqrt{\rho^{1/2}\sigma\rho^{1/2}}\right)^2$. Optimization of the fidelity over all input states on the Bloch sphere results in the qubit optimal universal cloner, in which $F=5/6$ \cite{UC0,UC1}. For orbital states, where $\theta$ is an \textit{a priori} known constant and $\phi\in[0,2\pi)$, a class of phase-covariant cloning machines has been suggested \cite{PCC2}. After the cloning process, in the computational basis $\{|0\rangle,|1\rangle\}$ (the eigenvectors of $\sigma_z=\text{diag}(1,-1)$) each of the clones can be identified by the density operator: $\varrho_{00}^\text{out}=\mu^2\varrho^\text{in}_{00}+\nu^2$ and $\varrho_{01}^ \text{out}=2\mu\nu\varrho^\text{in}_{01}$, where $\mu^2+2\nu^2=1$, and $\nu^2=1/6$ for UC and $\nu^2=(1-\frac{1}{\sqrt{1+2\tan^4 \theta}})/4$ for PCC. Most of this description is also valid when the original quantum system is initially mixed.

Our main assumption is that preparation of the initial pure state $|\Psi\rangle$ is diluted by a thermal bath in the following special and simple
form: \be&\varrho^\text{in}=(1-\epsilon)|\Psi\rangle\langle\Psi|+\epsilon\varrho_\text{th},\quad 0\leqslant\epsilon<1.\label{dilution}\ee The parameter $\epsilon$, which measures how thermally perturbed the preparation is, may in general be time-dependent. Nonetheless, based on our earlier assumptions, it would be a fairly slow-varying time-dependent function so that with a good approximation we can take it a relatively small constant of the order of $\tau_\text{c}/T_\text{diss.}$. This state does not seem to arise naturally from a typical thermalization dynamics. Nevertheless, in Ref.~\cite{BP} it has been illustrated that general behaviors obtained from such a simple preparation assumption (in the context of the geometric phases) have general features similar to those obtained from the Lindblad equation for the dynamics. It is worth mentioning that in the limit of infinite temperature, the
thermalized density matrix $\rho_\text{th}$ is equivalent to pure noise~\cite{wnoise}. In that case, $\epsilon$ represents the degree of pure noise existing during the process (
for example, in the case of NMR systems, due to fluctuations of the external magnetic fields and similar reasons). A more general analysis of quantum cloning in the presence of a thermalization mechanism is yet lacking, but our simple analysis may also shed some light before having a more complete analysis at hand.

\begin{figure}[t]
  \includegraphics[width=8.7cm,height=3cm]{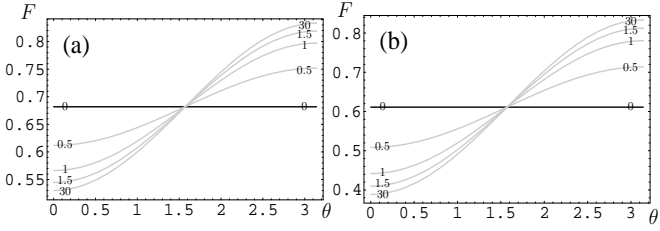}
  \caption{Fidelity ($F$) of UC vs $\theta$ for some values of $\eta$: (a) $\epsilon=5/11$ and (b) $\epsilon=2/3$.}\label{fig1}
\end{figure}

First, we consider the effect of the thermal term only on the state of the cloner, that is, the quantum cloning hardware is thermally diluted as in Eq.~(\ref{dilution}). In this case, the initial state of the machine is mixed. Considering the fact that in the optimal UC and PCC, the initial state of the cloning machine can be any pure state \cite{UC2,PCC1,blank}, one can conclude the optimal fidelity here is achieved by the existing optimal cloning transformations. By a similar analysis, it appears that for the case of diluted joint blank and ancillary systems, one can consider the joint state as a new blank copy and attach some new reservoir to the whole Hilbert space of the input states (i.e., the information qubit, the blank copy, and the ancilla state) as a new ancillary system and then define a new transformation for cloning \cite{blank}. This would in fact be the existing optimal cloning transformation, now acting on a larger Hilbert space, and hence one obtains the same optimal fidelity again. However, from an experimental point of view, thermalization effects are likely to occur during the cloning process rather than at the initial preparation level --- for instance in NMR systems~\cite{NMR1,NMR2}. Therefore, to be more precise, thermal effects during the preparation should also be taken into account.
\begin{figure}[t]
  \includegraphics[width=8.7cm,height=3cm]{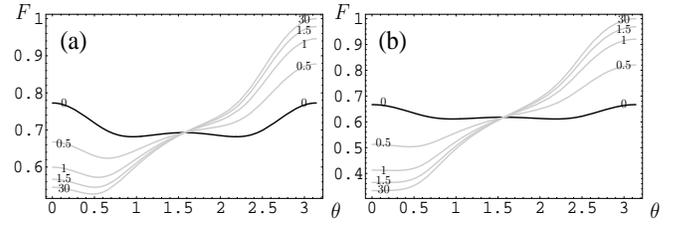}
  \caption{Variation in the fidelity of PCC with $\theta$ for: (a) $\epsilon=5/11$ and (b) $\epsilon=2/3$, and different values of $\eta$.}\label{fig2}
\end{figure}

\begin{figure}[t]
  \includegraphics[width=8.4cm,height=3.2cm]{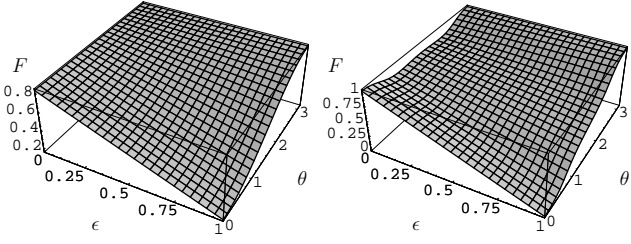}
  \caption{Fidelity vs $\epsilon$ and $\theta$ in low temperature limit ($\eta\rightarrow\infty$): UC (left) and PCC (right).}\label{fig3}
\end{figure}

Now, we consider the case in which the input state $|\Psi\rangle$ is thermally diluted as in Eq.~(\ref{dilution}). Our aim now is to compare the similarity between the clones and the input state of our interest, i.e., $|\Psi\rangle$. Indeed, here we assume that the model of the cloning machine consists of two parts: the first is the dilution of the input pure state which models the imperfect feature of the machine, and the second is some known UC or PCC transformation which is applied to this diluted state. The Hamiltonian of the qubit system is taken to be $H=\omega_0\sigma_z/2$ ($\omega_0>0$), whence, $Z=2\cosh\eta$, where $\eta=\omega_0\beta/2$. More general cloning transformations in spin networks with more complicated Hamiltonians can be found, for example, in Ref.~\cite{starnet}. The fidelity of the output state and the unperturbed initial state can be calculated as follows:\be\aligned F(\theta,\epsilon,\eta)=&\mu^2[1-\epsilon+\epsilon(e^{-\eta}\cos^2\frac{\theta}{2}+e^\eta\sin^2\frac{\theta}{2})/Z]\nonumber\\ &+(\mu\nu-\mu^2/2)(1-\epsilon)\sin^2\theta+\nu^2.\label{fidelity}\endaligned\ee Figure~\ref{fig1} illustrates how the fidelity in the UC behaves in terms of $\theta$ (orbit of the state) in thermally diluted states, for two different values of $\epsilon$ (the degree of thermalization) and $\eta~(\propto 1/T)$. It can be seen that when \be\epsilon<\cosh\eta~/(e^{-\eta}\sin^2\frac{\theta}{2}+e^\eta\cos^2\frac{\theta}{2}),\label{cond1}\ee the fidelity of the UC is higher than the classical value $1/2$. This threshold is the fidelity of a classical-like $1\to M$ universal cloning in which with a given probability, an unknown input state is sent to one of the $M$ parties and a completely randomized state is transmitted to any of the other ones, of course, in the limit of large $M$ \cite{CC1}. In the literature, however, ``classical cloner" has been attributed to some other cloning transformations as well --- see \cite{UC1,CC2}. In other words, in some cases thermal noise (even in the simple form of Eq.~(\ref{dilution})) can result in a lower performance than a classical machine. For $\theta\geqslant\pi/2$, the condition (\ref{cond1}) implies that for all $0\leqslant\epsilon<1$, the fidelity of the output of the UC is always greater than that of the classical cloner (if $\omega_0$ was negative, this would occur for $\theta\leqslant\pi/2$). Equation~(\ref{cond1}) can also be interpreted as a condition on temperature for a given $\theta$ and $\epsilon$ in order to outperform a classical cloner. Figure~\ref{fig2} shows the variation of the fidelity of the outputs of the PCC machines in terms of $\theta$, for some fixed values of $\epsilon$ and $\eta$. As is clear from this figure, in the case of equatorial qubits, similar to the case of the UC, the fidelity of the outputs does not vary with temperature --- according to Eq.~(\ref{fidelity}), this feature is due to the symmetry property of such states. Low temperature limits of the fidelity for both UC and PCC have been depicted in Fig.~\ref{fig3}. In the case of the UC, for all $\theta$ in $[0,\pi)$, the fidelity is a decreasing function of $\epsilon$. The corresponding graph for the PCC also shows a decrease in the fidelity for different values of $\theta\in[0,\pi/2)$ with the perturbation factor $\epsilon$. However, a closer inspection shows that here there are also some $\theta$s ($\gtrsim 2.52$ and less than $\pi$ rad) in which the fidelity of the PCC is an increasing function of $\epsilon$. At high temperature limit, the fidelity of both UC and PCC, for all $\theta$s, is a decreasing function of $\epsilon$. Another important point that can be concluded from the figures is that in some cases, the quality of the clones at the output of the UC can be better than that of the PCC --- see for example those regions of Fig.~\ref{fig3} in which $\epsilon$ and $\theta$ are large and small, respectively. This is indeed contrary to what happens when the cloning is performed perfectly without any external noise.
\begingroup
\squeezetable
\begin{table}[b]
\begin{ruledtabular}
\caption{Inseparability conditions of the output states in the three different scenarios of cloning.}
\begin{tabular}{cccc}
$\gamma$ & \multicolumn{2}{c}{$\alpha$} & $\epsilon$\\
\colrule\multicolumn{1}{c}{\multirow{3}*{$\gamma>\gamma_{\text{c}}$}} & \multicolumn{1}{|c}{\multirow{2}*
{\hskip1.5mm$|\alpha^2-1/2|<\alpha_{\text{c}}$}} & \multicolumn{1}{|c|}{$0<\alpha<1$} & $0\leqslant\epsilon<1$\\
\cline{3-4} & \multicolumn{1}{|c}{} & \multicolumn{1}{|c|}{$-1<\alpha<0$} & $0\leqslant\epsilon<\epsilon_1$ or
$\epsilon_2<\epsilon<1$\\
\cline{2-4} & \multicolumn{2}{|c|}{$|\alpha^2-1/2|\geqslant\alpha_{\text{c}}$} & $\epsilon_2<\epsilon<1$\\
\colrule\multicolumn{1}{c}{\multirow{2}*{$0<\gamma\leqslant\gamma_{\text{c}}$}} & \multicolumn{1}{|c}{\multirow{2}*
{\hskip1.5mm $|\alpha^2-1/2|<\alpha_{\text{c}}$}} & \multicolumn{1}{|c|}{$0<\alpha<1$} & $0\leqslant\epsilon<\epsilon_2$\\
\cline{3-4} & \multicolumn{1}{|c}{} & \multicolumn{1}{|c|}{$-1<\alpha<0$} & $0\leqslant\epsilon<\epsilon_1$\\
\end{tabular}
\label{table1}
\end{ruledtabular}
\end{table}
\endgroup

\emph{Entanglement cloning.}--- Quantum cloning can be used to clone or broadcast entanglement as well \cite{UC3,EC,Locc,LC,BK99}. Let us assume that we have an initial state in the form of $|\Psi^{-}_\alpha\rangle_{ab}=\alpha|01\rangle_{ab}-\sqrt{1-\alpha^2}|10\rangle_{ab}$, where $\alpha$ is real and $|\alpha|\leqslant1$. As in the cases of the UC and the PCC, suppose that because of a thermal environment, the initialization is diluted as in Eq.~(\ref{dilution}). Let us take our system to be two spin-1/2 particles interacting via the XX Hamiltonian: $H=J(\sigma_x^a\sigma_x^b+\sigma_y^a \sigma_y^b)$, where $\sigma_x$ and $\sigma_y$ are Pauli matrices. Now, we want to compare performances of the following schemes of entanglement broadcasting between two parties in the presence of thermal noise: (i) Local cloning by means of two optimal UC machines copying each qubit separately \cite{LC}. In this scenario, after the cloning process and discarding the ancillas, we will have the overall state $\varrho_{aa'bb'}$ whose two first (last) qubits are the copies of $a\,(b)$. (ii) Non-local cloning of the two-qubit state as a whole with the UC machine of $4$-level quantum states \cite{UC3}. (iii) Cloning by an optimal entanglement cloner \cite{EC}.
\begingroup
\squeezetable
\begin{table}[b]
\begin{ruledtabular}
\caption{Inseparability conditions of the output states in the three different scenarios of cloning, at low and high temperature limits.}
\begin{tabular}{cc}
\hskip8.5mm $\gamma$ & $\epsilon,\alpha$\\
\colrule \multicolumn{1}{c}{\multirow{5}*{ \hskip8.5mm $\gamma\rightarrow\infty$}} & \multicolumn{1}{|c}
{C1 \text{and} $0\leqslant\epsilon\leqslant\frac{1-M}{2M}$ \text{and} $|\alpha^2-1/2|<\alpha_1^\infty$}\\
\cline{2-2} & \multicolumn{1}{|c}{\hskip1mm C1 \text{and} $\frac{1-M}{2M}<\epsilon<1$ \text{and} $\alpha\in$ C1}\\
\cline{2-2} & \multicolumn{1}{|c}{\hskip1mm C2 \text{and} $0\leqslant\epsilon<\frac{3M-1}{4M}$ \text{and} $|\alpha^2-1/2|
<\alpha_2^\infty$}\\
\cline{2-2} & \multicolumn{1}{|c}{\hskip1mm C2 \text{and} $\frac{1-M}{2M}<\epsilon\leqslant\frac{M+1}{4M}$ \text{and}
$|\alpha^2-1/2|>\alpha_1^\infty$}\\
\cline{2-2} & \multicolumn{1}{|c}{\hskip1mm C2 \text{and} $\frac{M+1}{4M}<\epsilon<1$ \text{and} $\alpha\in$ C2}\\
\colrule \hskip8.5mm $\gamma\rightarrow0$ & \multicolumn{1}{|c}{\hskip1mm $0\leqslant\epsilon<(1-\frac{1}{3M})$
\text{and} $|\alpha^2-1/2|<\alpha^0$}\\
\end{tabular}
\label{table2}
\end{ruledtabular}
\end{table}
\endgroup

After some algebra, it can be seen that the density matrices of the clones in cases (ii) and (iii), and $\varrho_{a'b}$ (also $\varrho_{ab'}$, $\varrho_{ab}$, and $\varrho_{a'b'}$) --- nonlocal copies --- in case (i), read as follows:\be& \varrho^\text{out}=(\frac{M\epsilon}{Z}+ \frac{1-M}{4})(|00\rangle\langle 00|+|11\rangle\langle 11|)\nonumber\\&+[M(\frac{1-\epsilon}{2}+\frac{\epsilon\cosh\gamma}{Z})+\frac{1-M} {4}+L(1-\epsilon)(2\alpha^2-1)]|01\rangle\langle 01|\nonumber\\&+[M(\frac{1-\epsilon}{2}+\frac{\epsilon\cosh\gamma}{Z})+\frac{1-M}{4} -L(1-\epsilon)(2\alpha^2-1)]|10\rangle\langle 10|\nonumber\\&-M[(1-\epsilon)\alpha\sqrt{1-\alpha^2}+\frac{\epsilon}{Z}\sinh\gamma](|01\rangle\langle 10|+|10\rangle\langle 01|),~\ee in which $\gamma=2\beta J,\;Z=2(1+\cosh\gamma),\;{L}=3(1+2M+{\sqrt{1+4M-9M^2}})/26,\;M_\text{i}=(2/3)^2,\;M_\text{ii} =3/5,\;M_\text{iii}=6A^2+4AC$, $A=\sqrt{(1/2+1/\sqrt{13})}/3$, and $C=A(\sqrt{13}-3)/2$. Note that, the output states of case (ii) for all values of $\epsilon$, $\alpha$, and $\gamma$, the nonlocal copies of case (i) $\varrho_{a'b}$, and the output states of case (iii) for $\epsilon=1$ and $\forall\gamma$ or $\alpha=\pm 1/\sqrt{2}$ (for all $\epsilon$ and $\gamma$) all can be written in the following compact form: $\varrho^\text{out} =M\varrho^\text{in}+(1-M)I/4$, where $I$ is the $4\times4$ identity matrix.

\begin{figure}[t]
  \includegraphics[width=8.7cm,height=5.2cm]{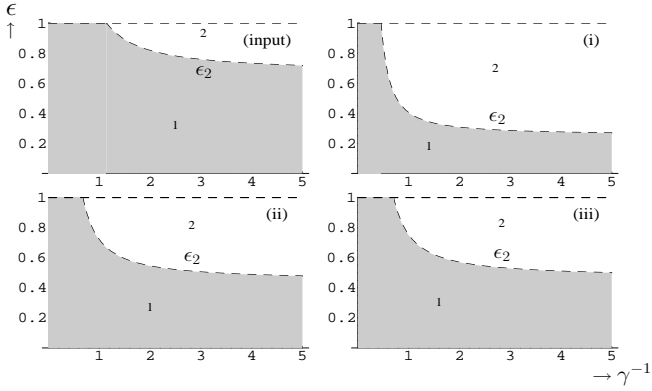}
  \caption{Entanglement phase diagrams of input and output states (achieved from three different schemes of entanglement cloning/broadcasting introduced in the text), when $\alpha=1/\sqrt{2}$. The regions labeled by 1 are the regions in which entanglement exists, whilst the regions
  labeled by 2 indicate no-entanglement regions. This figure shows that for $\gamma<\gamma_\text{c}$ ($T>T_\text{c}$), depending on the value of $\epsilon$, we may or may not have entanglement. $\gamma_c$ is a decreasing function of $M$. In other words, the area of region 1 increases when
  $M$ increases, as well. This may imply the advantage of the entanglement cloner $M_\text{iii}$ over the other entanglement broadcasting schemes.}\label{fig4}
\end{figure}

To determine the regions in which the output states are separable or inseparable, we use the well-known Peres-Horodecki positive partial transposition criterion \cite{PPT}. According to this criterion, in the case of $2\times 2$ and $2\times 3$ systems, a density matrix $\varrho_{AB}$ is inseparable (i.e., entangled) iff $(\varrho_{AB})^{T_A}$ ($T_A$: partial transposition with respect to system $A$) is not positive. Tables~\ref{table1} and \ref{table2} show the results for anti-ferromagnetic case ($J>0$). The parameters in the tables are as follows: \be\aligned&\alpha_1^\infty=\frac{ \sqrt{(3M-1)(M+1-4M\epsilon)}}{4M(1-\epsilon)},\\&\alpha_2^\infty=\frac{\sqrt{(M+1)(3M-1-4M\epsilon)}}{4M(1-\epsilon)},\\&\alpha_{\text{c}}= \frac{\sqrt{3M^2+2M-1}}{4M},\\&\gamma_{\text{c}}=\ln(\frac{M+1+2\sqrt{M^2+M}}{3M-1}),\\&\epsilon_{1(2)}=\frac{(M-1\mp4M\delta)(1+\cosh\gamma)} {2M[1\pm\sinh\gamma\mp2\delta(1+\cosh\gamma)]},\\&\alpha^0=\frac{\sqrt{\left(3M(1-\epsilon)-1\right)\left(M(1-\epsilon)+1\right)}}{4M(1-\epsilon)}, \endaligned\label{parameters}\ee where $\delta=\alpha\sqrt{1-\alpha^2}$, $\text {C}1\equiv0<\alpha\leqslant 1$, and $\text{C}2\equiv -1\leqslant \alpha\leqslant0$. When $\gamma\rightarrow\infty$ and $M=M_\text{iii}$, since $(3M-1)/4M>(1-M)/2M$, there exists an overlap between the $\epsilon$-inequalities in the third and fourth sub-rows of Table~\ref{table2}. In this case, one should notice that for $(1-M)/2M<\epsilon<(3M-1)/4M $, clones are entangled if $|\alpha^2-1/2| < \alpha_2^\infty$ or $|\alpha^2-1/2| > \alpha_1^\infty$. This removes the ambiguity in such cases.

Tables~\ref{table1} and \ref{table2} imply that in most temperature regions, the inseparability inequalities are not symmetric with respect to $\alpha\rightarrow -\alpha$. In other words --- unlike the case of $\epsilon=0$ --- depending on the sign of $\alpha$, the parameter regions over which the cloned pairs are entangled may be different. Another important point (see the second row of Table~\ref{table1}) is the existence of a critical temperature $T_\text{c}$ ($\propto 1/\gamma_\text{c}$) beyond which the cloned pairs for some $\alpha$ regions, $|\alpha^2-1/2|\geqslant \alpha_{\text{c}}$, for all $\epsilon$s are not entangled.

Overall, by taking into account the behaviors of the upper and lower bounds of the inseparability inequalities we can find that in some temperature regions, in Table~\ref{table1} (Table~\ref{table2}), there exist intervals of $\alpha^2\,(\epsilon)$ in which the cloned pairs are separable. The length of these intervals decreases when $M$ increases (recall that $M_\text{iii}>M_\text{ii}>M_\text{i}$). Furthermore, for a given $\alpha^2\, (\epsilon)$ at intermediate (two limits of) temperatures, the range of $\epsilon\,(\alpha^2)$ in which the clones are entangled increases when $M$ increases as well. Indeed, for some temperature regions, in Table~\ref{table1} (Table~\ref{table2}) there exist some $\alpha^2\,(\epsilon)$ in which clones for all $\epsilon\,(\alpha\text{ in C1 or C2})$ and all three $M$s are entangled --- e.g., see first sub-row of Table~\ref{table1} or second and fifth sub-rows of Table~\ref{table2}. These facts together with the entanglement phase diagrams in Fig.~\ref{fig4}, whose regions show existence of entanglement or its non-existence for $\alpha=1/\sqrt{2}$, indicate advantage of entanglement cloner $M_\text{iii}$, over the other cloning schemes. That is, the optimal entanglement cloner has an advantage over other mentioned schemes of entanglement broadcasting in the sense of robustness against thermal noise.

\emph{Conclusion.}--- We have studied the role of thermal noise in some quantum cloning schemes through a simple model of temperature effect on spin states at the input of the cloning machines. The performance of the cloning machines depends generally on the values of the thermal perturbation coefficient, the orbit of the original state on the Bloch sphere, as well as on the temperature. In addition, three scenarios of entanglement cloning of thermally diluted two-qubit states have been investigated. Our analysis shows that the clones generated from non-local transformations, in particular those out of the optimal entanglement cloner, remain entangled for wider regions of parameters. I.e., the optimal entanglement cloner shows a relatively larger region of entanglement in the parameter space. This can be considered as an advantage of optimal entanglement cloner over the other scenarios in the sense of robustness against thermal perturbations. This statement, however, is subject to the thermalization model we have used; so for a general conclusion a more detailed study is still needed. Our results may be of importance in practical implementations of quantum cloning in systems in which thermal effects are unavoidable, e.g., nuclear spin systems \cite{NMR1,NMR2}. Indeed, the large $\epsilon$ regime of our approach --- when $\tau_\text{c}$ is of the same order of magnitude as $T_\text{diss.}$ --- has already been experimentally realized in a different context \cite{NMR1}. This can be considered as a non-economic cloning process~\cite{FN}.

\emph{Acknowledgments.}--- Supports by the Center of Excellence in Complex Systems and Condensed Matter (CSCM) at Sharif University of Technology, \textit{i}CORE, MITACS, and PIMS are gratefully acknowledged.



\begin{thebibliography}{99}
\bibitem{FN} Other economic cloning transformations with small $\epsilon$ have also been analyzed in Ref.~\cite{NMR2}.
\end{thebibliography}
\end{document}